\documentclass[11pt,twoside]{article}
\usepackage{asp2010}

\resetcounters

\begin{document}

\title{On the nature of sub-millimetre galaxies}
\author{James S. Dunlop$^1$
\affil{$^1$Institute for Astronomy, University of Edinburgh, 
Royal Observatory, Edinburgh, EH9 3HJ, UK}}

\begin{abstract}
I discuss our current understanding of the nature of high-redshift ($z > 2$) 
(sub)-millimetre-selected galaxies, with a particular focus on whether their 
properties are representative of, or dramatically different from those 
displayed by the general star-forming galaxy population at these epochs. 
As a specific case study, 
I present some new results on the one sub-millimetre galaxy which happens 
to lie within the Hubble Ultra Deep Field and thus benefits from the very 
best available ultra-deep optical-infrared 
{\it Hubble Space Telescope}  and {\it Spitzer Space Telescope} imaging.
I then consider what these and other recent results from optical-infrared 
studies of sub-millimetre and millimetre selected galaxies 
imply about their typical 
masses, sizes and specific star-formation rates, 
and how these compare with those of other star-forming galaxies selected 
at similar redshifts.
I conclude with a brief discussion of the continued importance and 
promise of SCUBA2 in the era of {\it Herschel}.
\end{abstract}

\section{Introduction}
In recent years two alternative views of the high-redshift (sub)-millimetre 
selected galaxy population have emerged. Some authors have argued that the
galaxies uncovered via the deep SCUBA, MAMBO, AzTEC and LABOCA
sub-millimetre and millimetre wavelength 
surveys are basically high-redshift versions of the Ultra-Luminous 
Infrared Galaxies (ULIRGS) found in the local universe. In this picture 
they are thus 
galaxies of moderate 
stellar mass ($< 10^{11}\,{\rm M_{\odot}}$) involved in major mergers which 
produce a compact, short-lived starburst with a correspondingly extreme 
specific star-formation rate (SSFR=star-formation rate/galaxy stellar mass) 
(e.g. Gonzalez et al. 2011; Hainline et al. 2011; Engel et al. 2010). 
In contrast, others have 
argued (in some cases from the same primary data) that the evidence indicates 
that sub-millimetre galaxies simply represent 
the top end of the normal star-forming galaxy population at $z = 2 -3$. In 
this scenario they are galaxies of high stellar mass ($\simeq 1 - 4 
\times 10^{11}\,{\rm M_{\odot}}$), which still have sufficiently large 
reserves or supplies of cool gas to produce very high levels of star formation,
but in which this star formation is spatially-extended and is of a magnitude 
which is (at least on average) as ``expected'' 
given their large stellar masses and the typical SSFR at this epoch 
(e.g. Dav\'{e} et al. 2010, Targett et al. 2011; Rujopakam et al. 2011). 

The concept of a typical SSFR at $z > 2$ is a fairly new one, and 
comes from recent (still somewhat 
controversial) studies of the so called ``main sequence'' of star-forming 
galaxies which indicate that the actively star-forming galaxy population 
displays a characteristic SSFR (over a reasonably large range of mass)
which rises monotonically from $z \simeq 0$ to $z \simeq 2$ and appears to 
plateau at higher redshifts around a value of SSFR = $2 \pm 2\,{\rm Gyr^{-1}}$
(Noeske et a. 2007; Daddi et al. 2007; Stark et al. 2009; Gonzalez et al.
2010). This characteristic SSFR displayed by ultraviolet-selected and 
mid-infrared selected star-forming galaxies at $z > 2$ at least provides 
a helpful benchmark against which to judge whether or not sub-millimetre 
galaxies are genuinely pathological outliers from the general star-forming 
population in this epoch of maximum star-formation activity.

In this paper I briefly describe some of the latest observational evidence of
relevance to this debate, focussing first on some new results on the one 
sub-millimetre galaxy which happens to lie within the ultra-deep optical (ACS)
and near-infrared (WFC3/IR) imaging delivered by the 
{\it Hubble Space Telescope (HST)} in
the Hubble Ultra Deep Field (HUDF). This case study illustrates rather clearly
the challenge we continue to face in gaining a complete and robust 
understanding of high-redshift sub-millimetre galaxies. To properly establish
their bolometric luminosities, temperatures, star-formation rates,
gas/dust masses, stellar masses, pre-existing stellar populations, sizes and 
morphologies, requires the careful combination of the deepest available 
data spanning the wavelength range $\lambda \simeq 4000\,{\rm \AA} 
\rightarrow 20\,{\rm cm}$ which, with current facilities, means handling 
multi-frequency datasets which vary by a factor of up to $\simeq 500$ in 
angular resolution.

\begin{figure*}
\begin{center}
\includegraphics[width=0.6\textwidth]{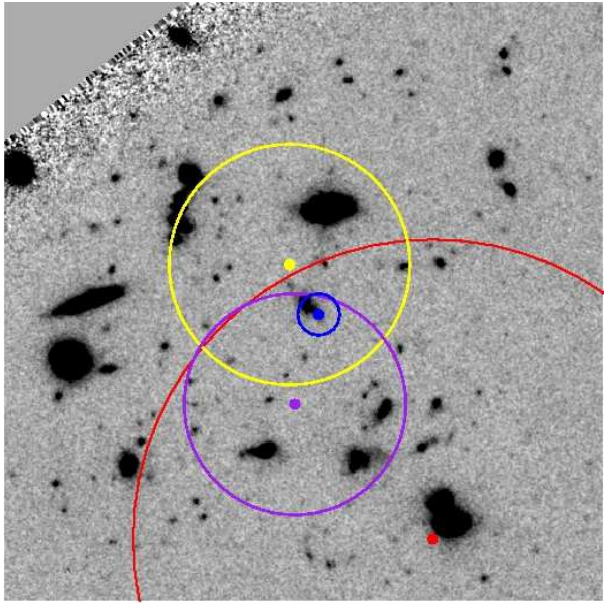}
\end{center}
\caption{The positions and appropriate counterpart search radii of the 3
alternative detections of the sub-millimetre/millimetre 
galaxy in the Hubble Ultra Deep Field (HUDF) are shown
superimposed on a greyscale of the appropriate sub-region of the
{\it HST} WFC3/IR $H_{160}$ image of the HUDF. The
position/search radius of the BLAST 250\,${\rm \mu m}$ detection is indicated
in red, the LABOCA 870\,${\rm \mu m}$ detection in purple, and the
AzTEC 1.1\,mm detection in yellow. Within the intersection of the three
counterpart search areas lies a single marginally-detected 1.4\,GHz radio
source which ties the object to a galaxy which is clearly detected
in the $H_{160}$ image, despite being almost undetected in the ACS $B$-band
image.}
\end{figure*}

\section{A luminous sub-millimetre galaxy in the Hubble Ultra Deep Field}
Surveys of the GOODS-South field at 870\,${\rm \mu m}$ with the 
sub-millimetre camera LABOCA (Weiss et al. 2009), at 1.1\,mm with the 
millimetre-wavelength camera AzTEC (Scott et al. 2010), 
and at 250--500\,${\rm \mu m}$ with the BLAST far-infrared imager 
(Devlin et al. 2009; Dunlop et al. 2010) 
have each detected one source which lies 
with the region of the HUDF 
recently imaged to unprecedented 
depth at near-infrared wavelengths with Wide Field Camera 3 (WFC3) 
on the {\it HST} (e.g. McLure et al. 2010; Bouwens et al. 2010). 
Despite the large beams, the three independent  
positions delivered by these millimetre--far-infrared facilities, 
combined with ultra-deep VLA imaging 
at 1.4\,GHz, have enabled the identification of a unique optical/infrared 
galaxy counterpart associated with the far-infrared dust emission, as shown 
in Figure 1. This galaxy is optically very faint ($B \simeq 30$\,AB mag), 
but also extremely red, 
and its location in the sub-region of the HUDF imaged by WFC3/IR provides a 
unique opportunity to study an ``ordinary'', unlensed sub-mm galaxy with 
{\it extra-ordinary} optical/infrared/mid-infrared imaging.
As shown in Figure 2, the unparalleled 12-band optical-infrared photometry 
delivers an extremely robust and accurate 
photometric redshift of $z = 2.97 \pm 0.09$.

\begin{figure*}
\includegraphics[width=1.0\textwidth]{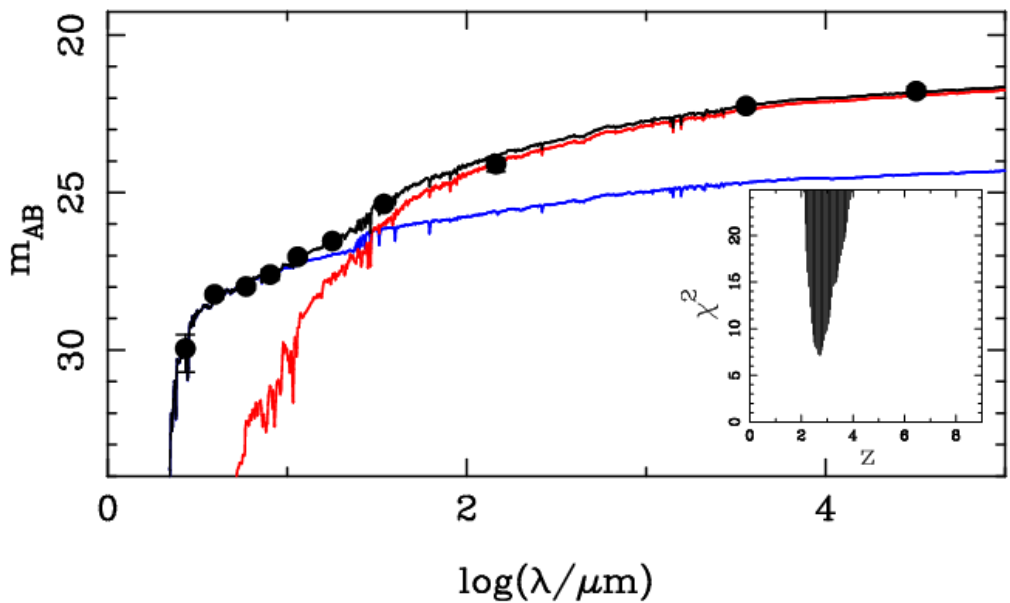}
\caption{The best-fitting two-component SED fit to the {\it HST}-ACS + 
{\it HST}-WFC3/IR + 
{\it Spitzer}-IRAC photometry of the galaxy associated with the 
sub-millimetre source
in the HUDF. The $\chi^2$ inset shows that the only acceptable solution is a 
galaxy with an accurate and robust photometric redshift $z = 2.97 \pm 0.09$.
The blue line indicates the contribution of the younger star-forming component
to the overall SED fit (shown in black), while the red line indicates 
the contribution of the mass-dominant more mature stellar population. The 
inferred stellar mass is $M_* \simeq 
2.5 \times 10^{11}\,{\rm M_{\odot}}$, with 
uncertainties dominated by the modelling choices as discussed in the text.}
\end{figure*}

\section{Stellar masses}

The stellar mass of this HUDF galaxy as derived from the two-component 
spectral-energy distribution (SED) 
fit shown in Figure 2 is $M_* \simeq 2.5 
\times 10^{11}\,{\rm M_{\odot}}$. This 
value is derived using the evolutionary synthesis models of Bruzual \& Charlot
(2003), and assuming the initial mass function (IMF) of Chabrier (2003). As 
has been well documented in the literature, the choice of evolutionary 
synthesis model, IMF, and also adopted star-formation history 
all have systematic effects on estimated 
galaxy stellar masses, and so it is important 
to quantify the impact of such choices on derived stellar masses, and to assess the 
evidence for or against the various alternatives.

The first potential cause of significant uncertainty in derived stellar masses 
is redshift accuracy, and it is undoubtedly the case that the 
{\it random} errors in stellar masses are in effect often dominated by 
uncertainty in estimated redshifts. Such uncertainty is obviously ideally 
removed via a spectroscopic redshift but, in the case of the HUDF galaxy 
considered here, such spectroscopy is both impractical but also not really
necessary, as the quality of the photometry yields such an unambiguous and 
accurate photometric redshift. Given high-quality photometry and a 
solid redshift, the uncertainty in stellar mass is undoubtedly 
dominated by choices made in the model-fitting process.

The first key modelling issue is the choice of star-formation history (SFH). 
The degree to which this is truly a free choice of course 
depends to some extent 
on the quality of the 
photometry. In an attempt to account for this uncertainty, 
Hainline et al. (2011) 
explored the use of two alternative single-component SFHs, adopting 
either an instantaneous starburst or a continuous star-formation history. 
Hainline et al. (2011) then averaged the two resulting estimates to produce the 
adopted stellar mass for each galaxy. In contrast, other authors
(e.g. Michalowski et al. 2010) have 
assumed a two-component SFH, while others have attempted to go 
further and fit several independent components (e.g. Dye et al. 2008).

The assumption of a multi-component SFH generally leads to higher 
mass-to-light ratios and, in turn, higher stellar masses than the
use of a single-component model. This is because while the 
starburst component can account for the ultraviolet (UV) emission 
(and the far-infrared emission
from the UV-heated dust), the second (older) component is then free to 
contribute more stars with higher mass-to-light ratios (see also 
Schael et al. 2011). By contrast, the use of a single instantaneous starburst 
model limits the age of the {\it entire} stellar population to the young age 
of the starburst required to match the UV emission 
(and thus the true stellar masses are inevitably under-estimated), 
while in the continuous star-formation model the current star-formation
rate is 
set by the UV emission and the total age is then limited in order not to 
overshoot the optical and the near-infrared part of the spectrum 
(assuming the galaxy has always formed stars at the same high rate).
Indeed, Schael et al. (2011) have found that the stellar masses of the 
SHADES sub-millimetre galaxies (Coppin et al. 2006) derived using a 
two-component SFH are on average a factor of $\simeq 2$ higher than when
derived with a one-component SFH. 

In the case of the HUDF sub-millimetre galaxy shown here in Figure 2, the 
situation is relatively straightforward because the 
optical-infrared data are of sufficient quality to demand a two-component fit
(but do not require anything more complex). This object thus provides some 
further support for the general adoption of two-component fitting when 
modelling the optical--mid-infrared SEDs of sub-millimetre galaxies, thus 
favouring higher stellar masses compared to those derived on the 
assumption of single-component star-formation histories.

The second key modelling issue is the choice of 
stellar population evolutionary 
synthesis model. Here the new HUDF sub-millimetre galaxy can provide little 
guidance, as satisfactory fits can be achieved with a variety 
of spectral synthesis models. However, the key issue for the derivation of 
stellar masses centres on whether to adopt models which have a strong 
contribution from thermally pulsating asymptotic giant branch (TP-AGB) stars
(as exemplified by the models of Maraston 2005), or models in which 
this phase of stellar evolution makes a relatively unimportant contribution
to the integrated light from a galaxy's stellar population (as is the case 
in the models of Bruzual \& Charlot 2003). Hainline et al. (2011) explored
both alternatives, and found that the stellar masses derived using the 
Bruzual \& Charlot (2003) models were on average $\simeq 50$\% higher 
than those calculated using the Maraston (2005) models. 

While studies of sub-millimetre galaxies may not be able to directly resolve this issue, 
several other recent studies of other classes of galaxy indicate that the 
impact of the TP-AGB stars on integrated galaxy light is not nearly as 
strong as produced by the Maraston (2005) models. In particular, 
Kriek et al. (2010) have found that the average spectral energy distribution 
of post-starburst galaxies at $0.7 < z < 2.0$ does not
appear to display the excess near-infrared contribution predicted from the 
TP-AGB contribution in the Maraston models, and that the Bruzual \& Charlot 
(2003) model provides a much better description of the data, 
even at the age ($\simeq 1$\,Gyr) 
when the TP-AGB contribution should be near maximum (see 
also Conroy \& Gunn 2010). For now, therefore, 
we favour the use of Bruzual \& Charlot
(2003) models over the Maraston (2005) 
models for the estimation of stellar masses.

The third analysis choice of importance for stellar mass calculation
concerns the adopted stellar IMF. The choice clearly matters because the 
adoption of a Salpeter (1955) IMF yields stellar 
masses a factor of $1.8$ higher 
than result from the adoption of the IMF favoured by Chabrier (2003).
In general, the discussion of the IMF in the context of sub-millimetre 
galaxies has until recently focussed on whether a top-heavy 
IMF is required to reproduce the sub-millimetre number counts from 
current models of galaxy evolution 
(e.g. Baugh et al. 2005; Fontanot et al. 2007; Hayward et al. 2011).
However, recently van Dokkum \& Conroy (2010) have 
reported that the IMF in  
local ellipticals (the likely descendants of sub-millimetre galaxies) 
is, if anything, slightly steeper
(bottom-heavy) than even the Salpeter (1955) IMF (see also 
Thomson \& Chary 2011). Given this controversy, and 
since there is as yet no clear evidence that the 
IMF of sub-millimetre galaxies is
systematically different from that of other galaxies, it makes sense to adopt 
the Chabrier (2003) IMF. This IMF was derived for the Milky Way and 
can be viewed as a conservative/neutral choice, in 
effect representing the middle ground between 
the top-heavy IMFs proposed by, for example, Baugh et al. (2005), and 
the bottom-heavy IMFs of Salpeter (1955) or van Dokkum \& Conroy (2010).
I also note that this ``Chabrier IMF'' is essentially
identical to the ``Canonical IMF'' recently summarized by 
Weidner, Kroupa \& Pflamm-Altenburg (2011). 
While I would argue that this IMF is the natural choice
on the basis of current evidence, it must still be accepted that we have to 
live 
with a fundamental uncertainty of $\sim \times 2$ in stellar masses until
the uncertainty in the IMF in sub-millimetre galaxies is resolved. 
However, it is also important to 
note that the derived specific star-formation rate (SSFR) is,
to first order, unaffected
by this IMF uncertainty (it is of course possible that the observed starburst 
and the pre-existing mass-dominant stellar population have different IMFs), making it a 
particularly useful quantity for comparing the 
properties of different types of 
star-forming galaxies, as discussed further below.

In conclusion, the above discussion hopefully clarifies why our current 
preferred best estimate of the stellar mass of the sub-millimetre galaxy 
in the HUDF is based on a two-component SED fit, using Bruzual \& Charlot 
(2003) models, and assuming a Chabrier (2003) IMF. 
The resulting fairly large stellar mass of 
$M_* \simeq 2.5 \times 10^{11}\,{\rm M_{\odot}}$ is 
consistent with the results reported by Schael et al. 
(2011) and Michalowski et al. 
(2010). Such values are substantially 
larger than the typical stellar masses of 
$M_* \simeq 5-7 \times 10^{10}\,{\rm M_{\odot}}$
reported by Hainline et al. (2011). 
However, they are not as extreme as some of the 
very large masses reported by Dye et al. (2008) 
using multi-component fitting and a 
Salpeter (1955) IMF.

\section{Dynamical and gas masses}

We do not currently possess any CO spectroscopy of the 
sub-millimetre galaxy in the HUDF,
but such data have now been gathered and 
analysed for a number of sub-millimetre 
galaxies, providing vital information on 
both the dynamical and molecular $H_2$ gas 
masses in these objects (e.g. Tacconi et al. 2006; 2008). 
Such measurements can be used to provide a consistency test of the stellar 
mass estimates, and to infer the evolutionary state of the galaxy.
Specifically, assuming that the extent of the CO line-emitting gas 
is the same as the extent of the stellar component, the sum of 
the gas and stellar masses should, within the uncertainties, be no larger than the 
dynamical mass. Using this approach, Engel et al. (2010) 
concluded that the stellar masses of the sub-millimetre galaxies derived by 
Michalowski et al. (2010)
($\simeq 2-3 \times 10^{11}\,{\rm M_{\odot}}$ after correction to the Chabrier (2003) IMF) 
are inconsistent with their gas and dynamical masses ($M_{\rm dyn} \simeq 3 \pm 1 \times
10^{11}\,{\rm M_{\odot}}$; see also Tacconi et al. 2008). 
However, not only does this result appear to be statistically insignificant, 
but on the contrary it appears that new observations of CO 1-0 line emission
mean that this test now in fact appears to provide further support for 
large stellar masses $M_* > 10^{11}\,{\rm M_{\odot}}$.

Like most previous CO line observations of high-redshift 
sub-millimetre galaxies,
Engel et al. (2010) detected the high-excitation 
lines (CO J=7-6, 6-5, 4-3 and 3-2), 
which trace denser and hotter gas, arguably likely confined to the central regions 
of a galaxy. Therefore, the derived dynamical masses may refer to regions smaller 
than the extent of the stellar component, so it is difficult to compare them 
directly with the stellar masses, given that the starlight typically has a 
half-light radius of $r_{0.5} 
\simeq 3-4$\,kpc (see below). Indeed, based on new 
VLA studies of low excitation CO(1-0) lines, Ivison et al.
(2011) have found dynamical masses higher by a factor $\simeq 2.4-4.1$ 
for two out of the three sub-millimetre galaxies for which Engel et al. (2010)
have claimed inconsistencies between stellar and dynamical masses
(HDF 76 and N2 850.2; the third one was not observed by Ivison et al. 2011).
It is also worth noting that the mean value of $M_{\rm dyn}$ 
reported by Engel et al. (2010) is lower by a factor of $\simeq 1.7$ (close to the
magnitude of the claimed inconsistency in the stellar masses of Michalowski et al.
2010) than that typically derived from near-infrared integral-field
spectroscopy of sub-millimetre galaxies 
($M_{\rm dyn} \simeq 5\pm3\times10^{11}\,{\rm M_\odot}$; Swinbank et al. 
2006), although both values are formally consistent.

Adopting the CO(1-0) derived masses where available, and multiplying masses based on
higher-frequency CO transitions by a factor of $2$ (as inferred from 
Ivison et al. 2011), 
we find that the average value of the ratio 
$(M_*+M_{\rm gas})/M_{\rm dyn}$ is $1.18\pm0.30$, adopting the stellar masses 
of Michalowski et al. (2010) after correcting to the Chabrier (2003) IMF.
We conclude that the latest estimates of (sub)-millimetre dynamical masses are 
in good accord with their stellar masses, and certainly cannot be used to rule 
out the higher values as inferred here from the adoption of 
two-component modelling and the 
Bruzual \& Charlot (2003) models. However they do provide some 
tentative evidence against
adoption of the Salpeter (1955) IMF which would push the inferred stellar masses 
of several (sub)-millimetre galaxies beyond the typical CO(1-0) inferred dynamical 
mass of $M_{\rm dyn} \simeq 5 \times 10^{11}\,{\rm M_{\odot}}$.

Calculation of molecular ($H_2$) gas masses in sub-millimetre galaxies 
is currently somewhat controversial due to uncertainty over what is the 
appropriate correction factor to apply when converting CO luminosities into 
total molecular gas masses. Here I simply note that given dynamical masses 
$M_{\rm dyn} \simeq 5 \times 10^{11}\,{\rm M_{\odot}}$ 
and typical stellar masses $M_* \simeq 3 \times 10^{11}\,{\rm M_{\odot}}$, current 
data would appear to allow molecular gas masses up to levels comparable with the 
stellar masses, which in fact would then allow CO luminosity to gas-mass conversion 
ratios possibly as large as the Milky Way value of $M_{gas}/L^{\prime}_{CO}
= \alpha_{CO} = 4.5\,{\rm M_{\odot}\,(K\, km\,s^{-1}\,pc^2)^{-1}}$, 
or certainly $\alpha_{CO} = 3.2 - 3.6\,{\rm M_{\odot}\,(K\, km\,s^{-1}\,pc^2)^{-1}}$ (comparable to the coversion factor implied for nearby disc galaxies; 
Tacconi et al. 2010), rather than necessarily 
requiring $\alpha_{CO} \simeq 0.8\,{\rm M_{\odot}\,(K\, km\,s^{-1}\,pc^2)^{\
-1}}$ as commonly adopted for ULIRGS (e.g. 
Downes \& Solomon 1998; Tacconi et al. 2006; Daddi et al. 2010).

\begin{figure*}
\begin{center}
\begin{tabular}{ccc}
\includegraphics[width=0.31\textwidth]{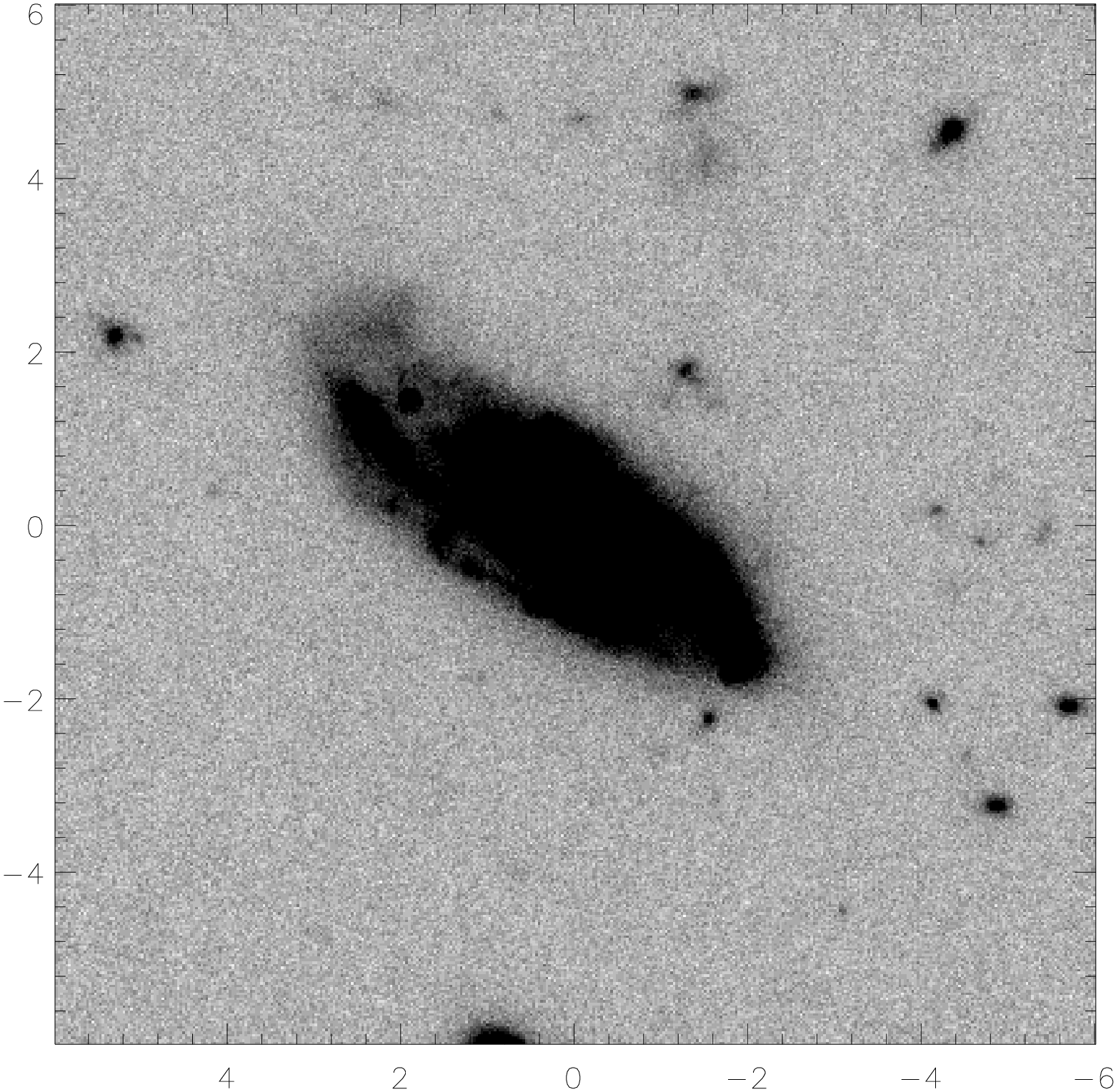}&
\includegraphics[width=0.31\textwidth]{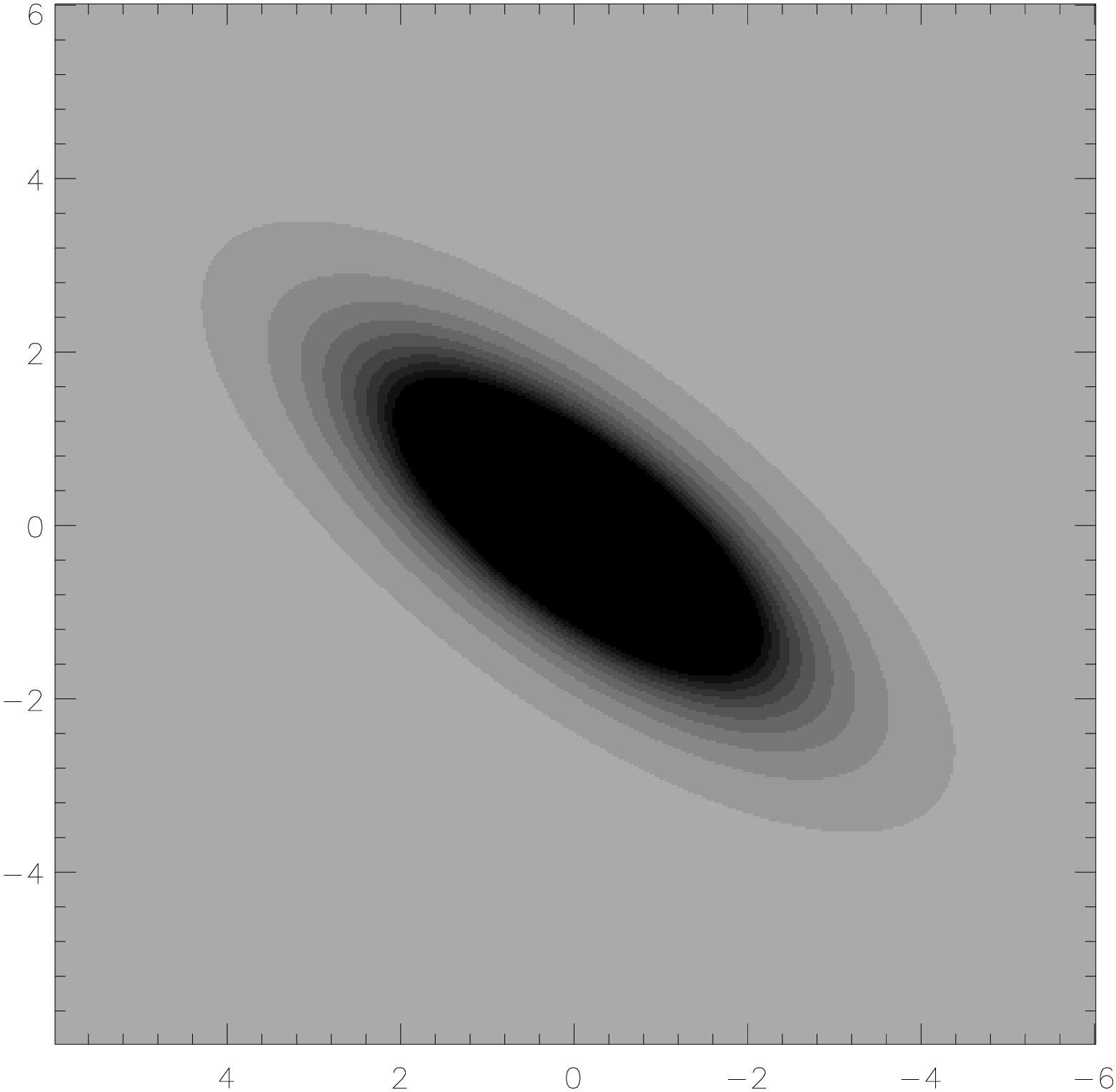}&
\includegraphics[width=0.31\textwidth]{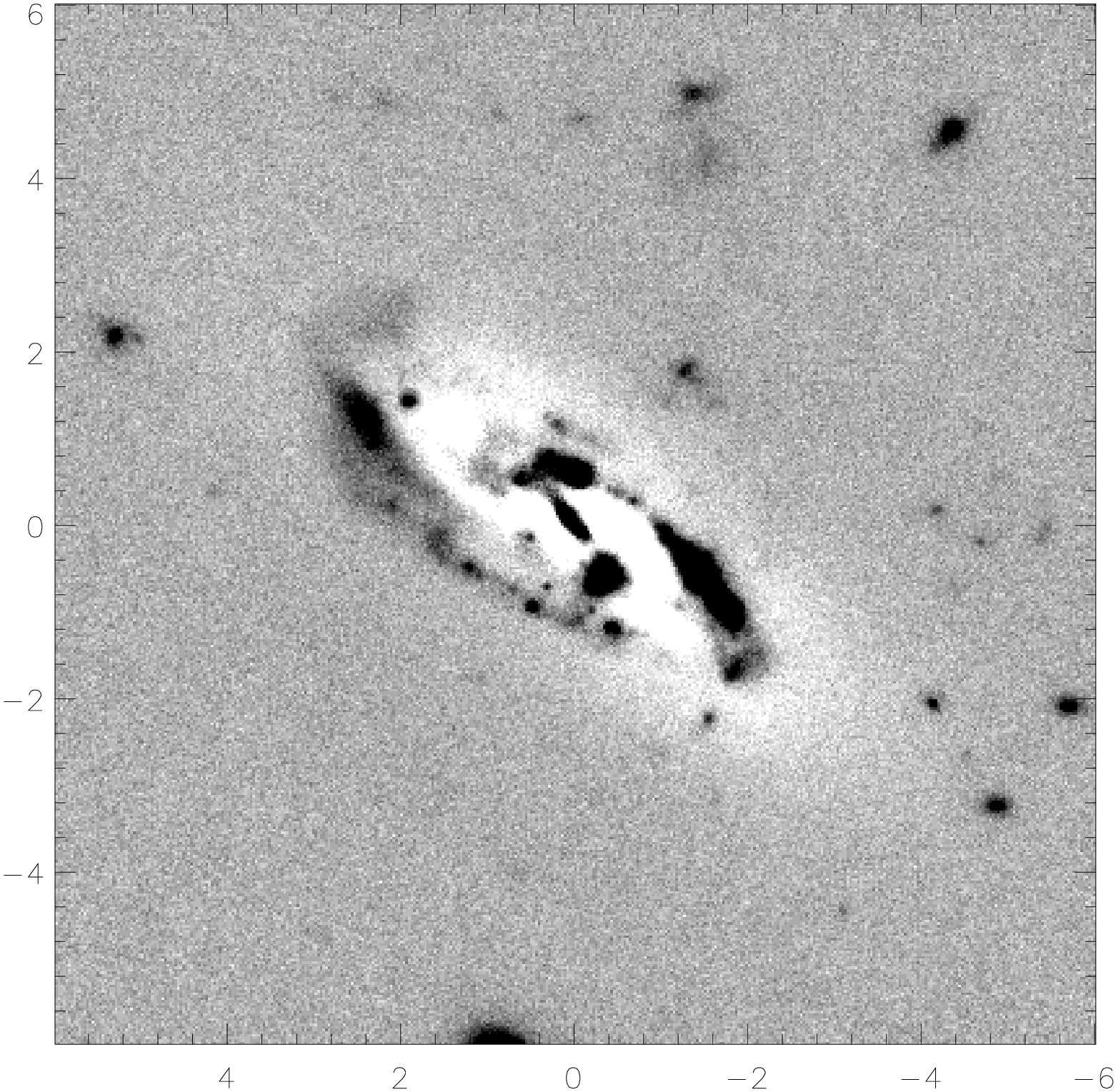}\\
\\
\includegraphics[width=0.31\textwidth]{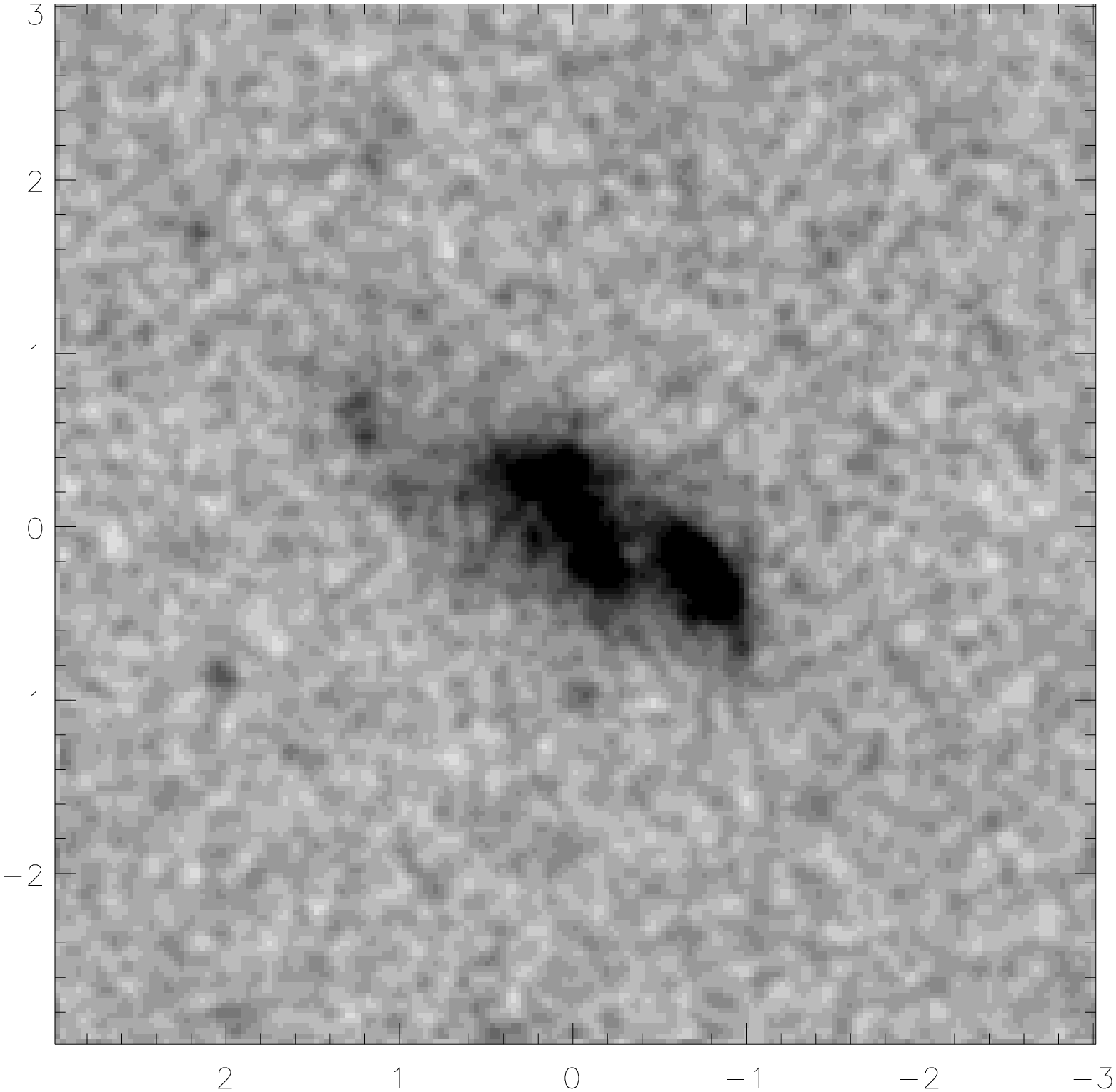}&
\includegraphics[width=0.31\textwidth]{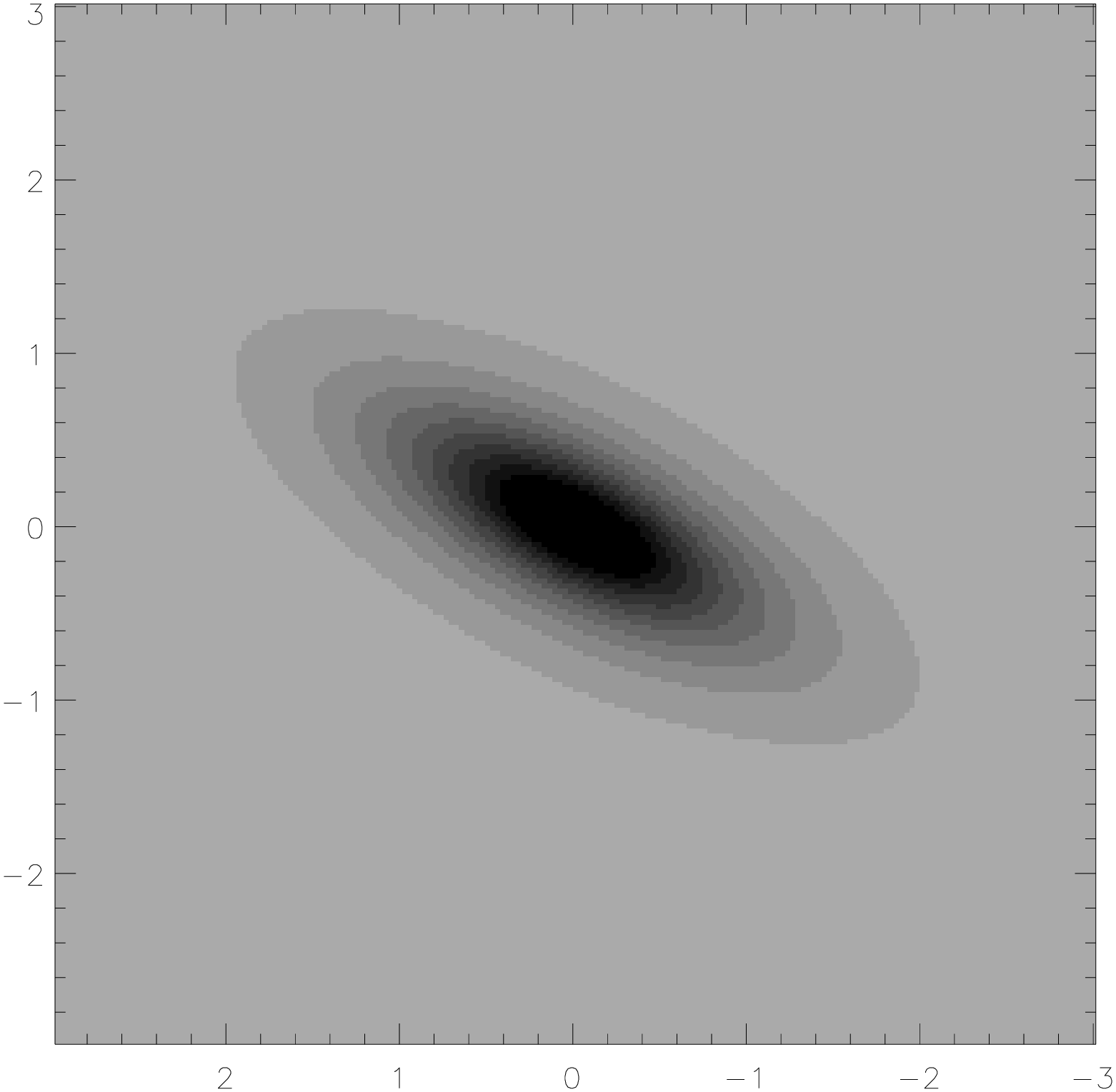}&
\includegraphics[width=0.31\textwidth]{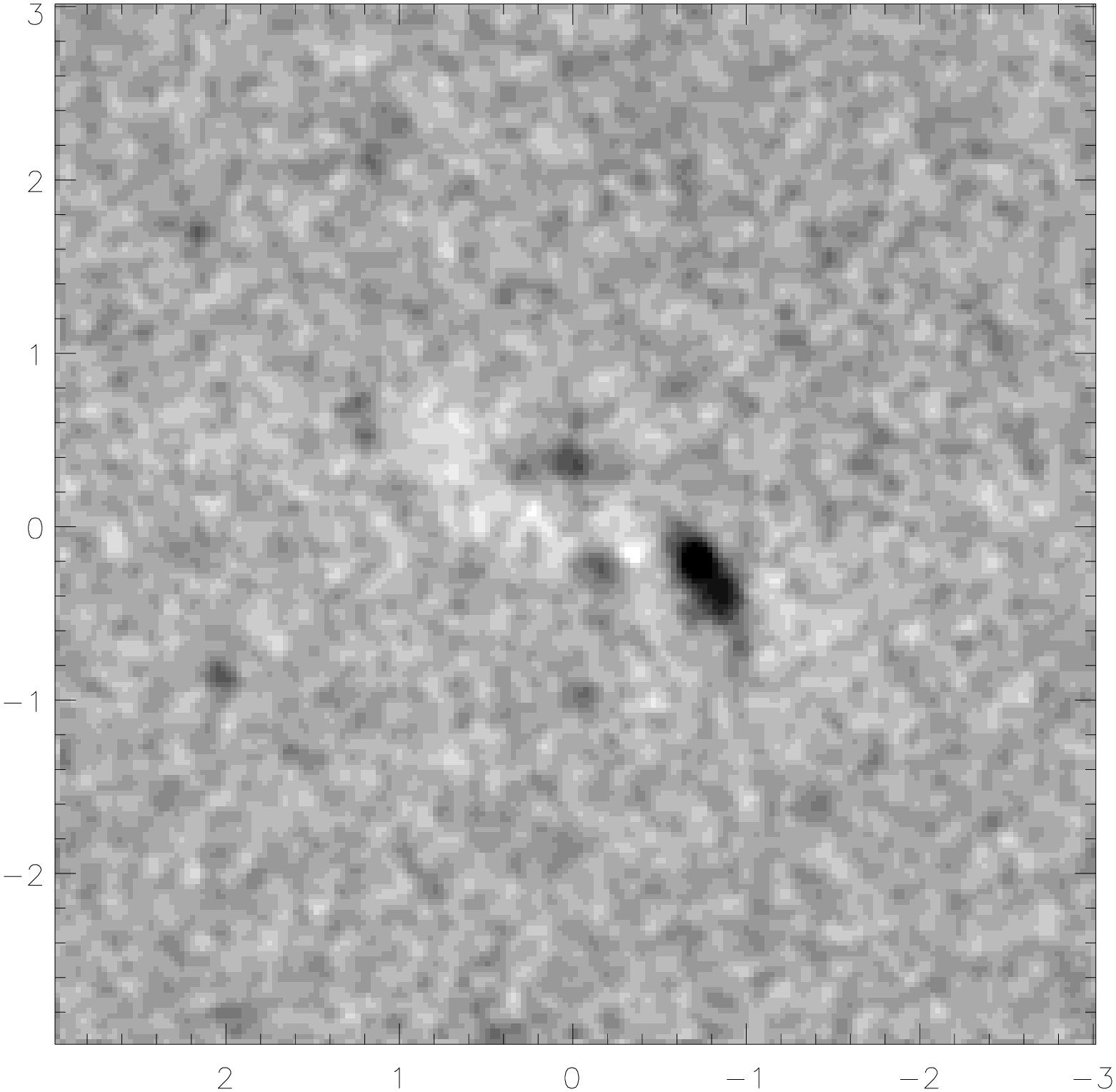}\\
\\
\includegraphics[width=0.31\textwidth]{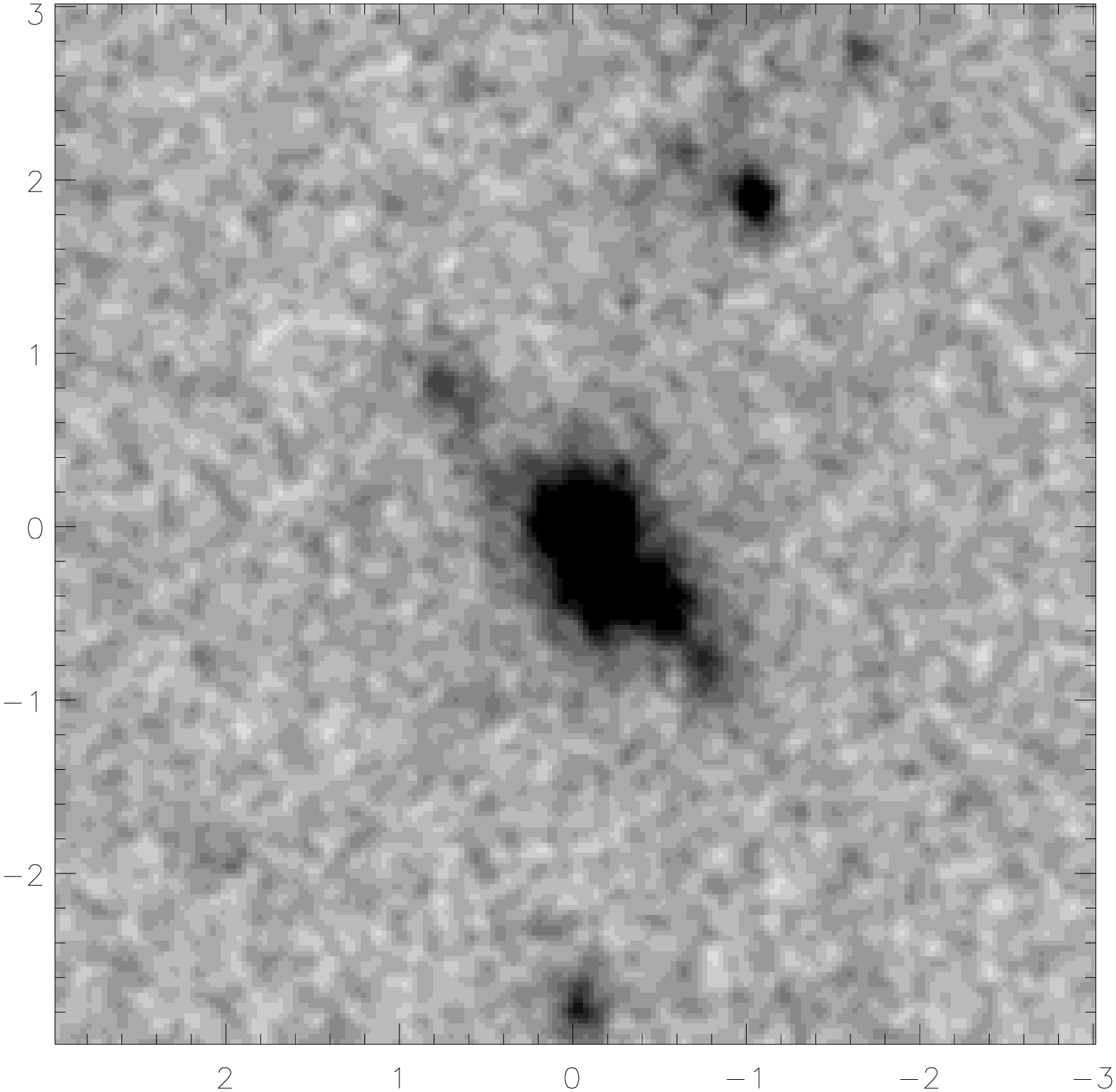}&
\includegraphics[width=0.31\textwidth]{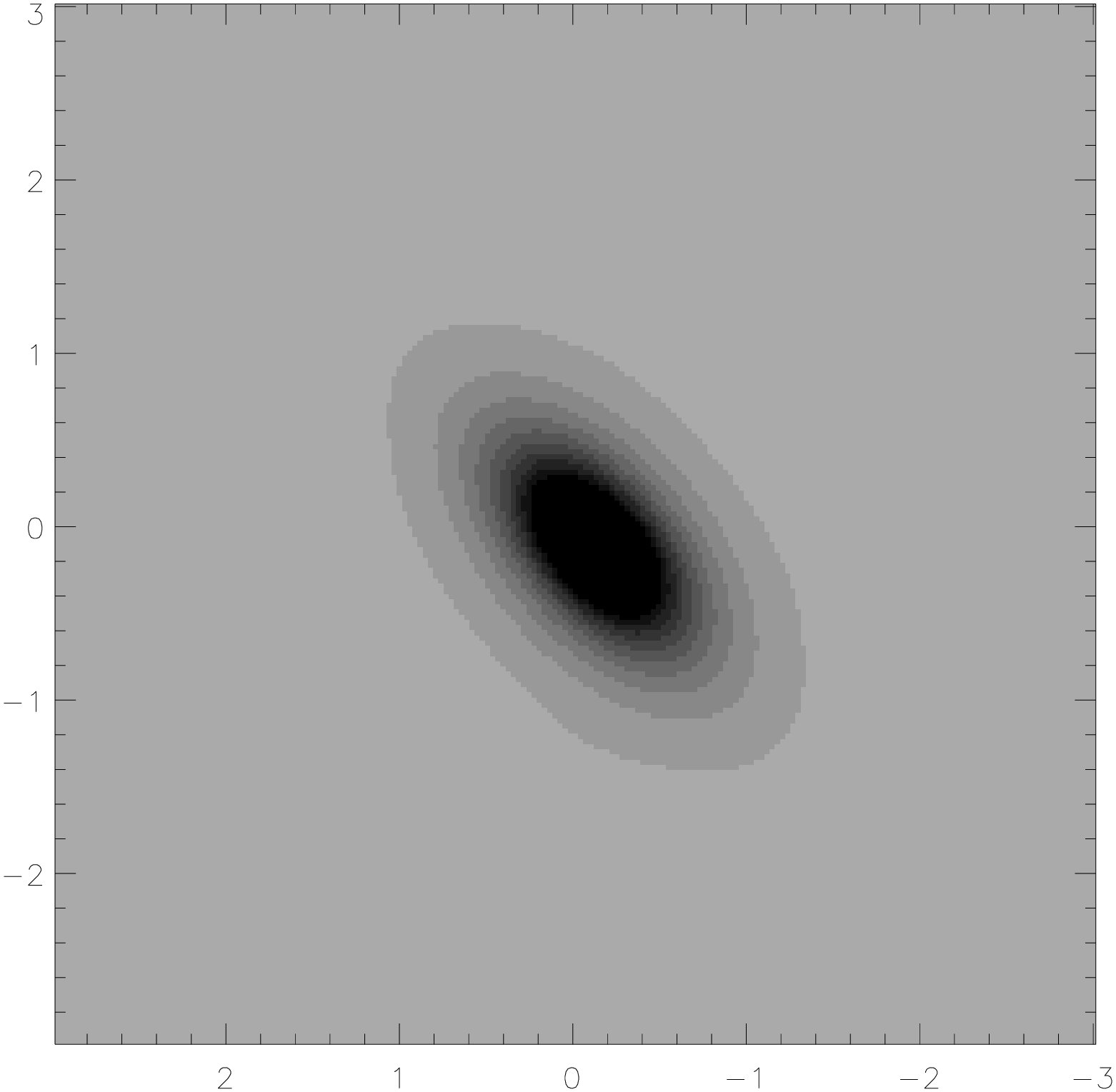}&
\includegraphics[width=0.31\textwidth]{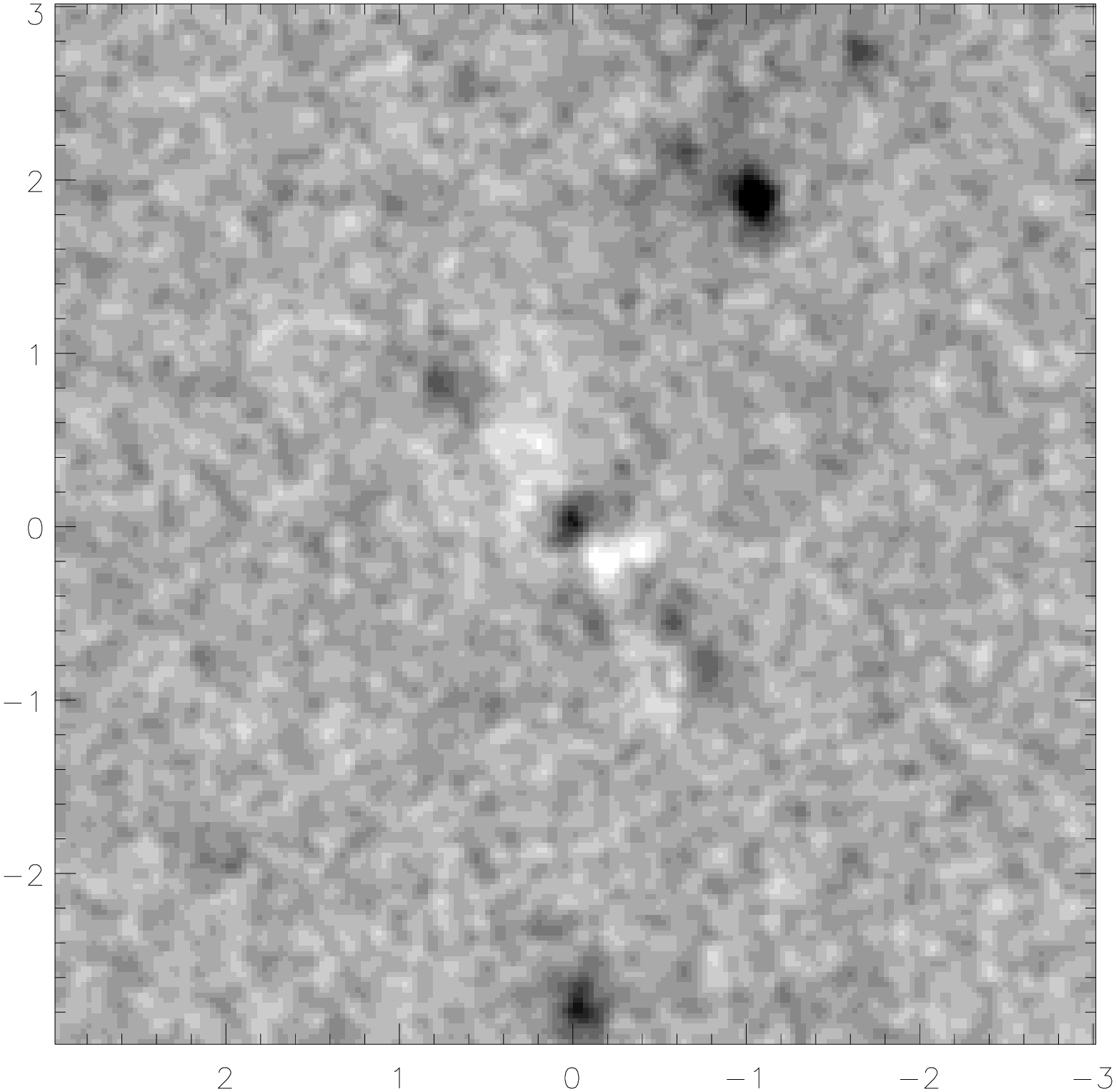}\\
\end{tabular}
\end{center}
\caption{An investigation into the rest-frame optical morphology 
of the $z = 3$ sub-millimetre galaxy 
in the HUDF, via two-dimensional modelling 
and comparison with a $z = 0.345$ spiral galaxy extracted from within the 
HUDF ACS optical imaging. The top row of 12 x 12 arcsec 
panels shows, from left to right, 
the $V$-band image of the $z = 0.345$ spiral, the best-fitting model, 
and the residual model-subtracted 
image (the best-fitting model is a disc galaxy 
with half-light radius $r_{0.5} = 8$\,kpc). The central 
row shows the effect of moving 
this galaxy to $z = 3$, as then imaged in the $H_{160}$-band with WFC3/IR. 
Again the modelling reclaims the correct physical scalelength, but now the 
residual image looks like a single ``clump''. The bottom row shows analogous 
plots for the actual HUDF sub-millimetre galaxy. The best-fitting model is
a disc galaxy (S\'{e}rsic index $n = 1.09$) with a half-light radius of $r_{0.5} = 4$\,kpc).}

\end{figure*}

\section{Sizes and morphologies}

Because of the exquisite depth and high angular resolution of the WFC3/IR 
imaging, it is possible to determine rather accurately the 
rest-frame ($B$-band) optical morphology of the (sub)-millimetre galaxy 
in the HUDF. 
As illustrated in Figure 3 (bottom row of panels) the result of fitting a 
two-dimensional axi-symmetric model with variable S\'{e}rsic index, is 
that the galaxy is a disc ($n = 1.09$) with a reasonably large half-light 
radius $r_{0.5} \simeq 4$\,kpc. These parameter values are similar to the
(albeit more uncertain) values derived using ground-based 
near-infrared imaging of $\simeq 15$ sub-millimetre galaxies by 
Targett et al. (2011), who reported median values of 
$n = 1.08$ and $r_{0.5} = 3.1$\,kpc. Similar sub-millimetre galaxy 
rest-frame optical scalelengths have also been 
reported from {\it HST} NICMOS imaging by Swinbank et al. (2010).
This case-study thus further strengthens 
the conclusion
that typical sub-millimetre 
selected galaxies are massive star-forming disc galaxies 
at $z \simeq 2 - 3 $.

To explore the reliability of this result, we have also analysed the 
HUDF imaging of a low-redshift ($z = 0.345$) disc/spiral galaxy, first
from the ACS imaging at its actual redshift (Figure 3, top row) and second
after artificially shifting it to $z = 3$, convolving it with the WFC3/IR 
$H_{160}$-band 
point-spread function, and planting the synthesized high-redshift 
spiral into the WFC3/IR $H_{160}$ image (Figure 3, middle row). For a 
single object, the results can of course 
only be regarded as illustrative, but 
nonetheless are worthy of comment. First, the modelling of the actual ACS 
optical imaging delivers a best-fit model of a disc galaxy with 
$r_{0.5} = 8$\,kpc, and the expected spiral arm residual. Second, when 
the synthesized $z = 3$ near-infrared image of the same galaxy is analysed, 
the model fitting returns identical parameters to within an accuracy of 
$\simeq 5$\%, providing reassurance that, with near-infrared imaging of this quality and depth,  the best-fitting parameters for the
HUDF (sub)-millimetre galaxy can certainly be regarded as robust. 
Third, the residual image 
from the modelling of the redshifted spiral is dominated by a single, 
off-centre clump. This is exactly the sort of feature that might naively 
be interpreted as evidence for an interaction/merger, but in fact it 
can be seen that this is simply a result of one of the brightest knots in the 
spiral arms coming to dominate the residual image after convolution 
with the WFC3/IR $H_{160}$ point-spread function. 

This experiment 
at least illustrates that one should be cautious in interpreting 
apparent secondary clumps in the near-infrared images of high-redshift 
sub-millimetre galaxies as evidence for interactions/mergers fueling or 
triggering the 
the extreme star-formation activity. In fact, as can been seen from the final 
panel in Figure 3, the actual sub-millimetre galaxy displays little obvious 
sign of a significant secondary component after subtraction of the 
best-fitting axi-symmetric disc-galaxy model. The high star-formation
rate in this galaxy could of course be merger-driven 
but, if so, then there is little obvious evidence of such 
an event in this, the very best available {\it HST} 
near-infrared imaging.

\begin{figure*}
\includegraphics[width=1.0\textwidth]{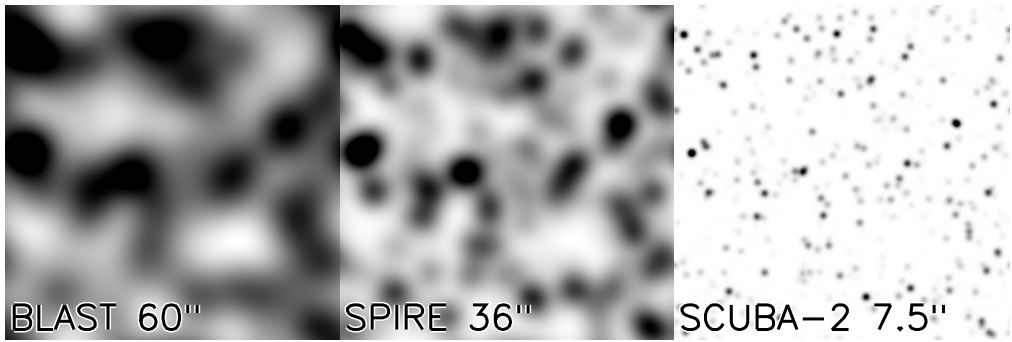}
\caption{A comparison of BLAST 500\,${\rm \mu m}$, {\it Herschel} SPIRE
500\,${\rm \mu m}$ and (predicted) SCUBA2 450\,${\rm \mu m}$ images of
a 50-arcmin$^2$ region of sky, demonstrating the potential power of
SCUBA2. A realisation drawn from BLAST 500\,${\rm \mu m}$ source counts
(extracted via P(D) analysis by Patanchon et al. 2009)
has been used to create the three 50 arcmin$^2$ simulated images shown in the
grey-scale panels, where the source population has been convolved with the
appropriate beam sizes for BLAST, {\it Herschel}+SPIRE and JCMT+SCUBA2 as
indicated in each panel (Devlin et al. 2009; Griffin et al. 2010; Holland
et al. 2006). The synthesized SCUBA2 image on the right
contains $\simeq 30$ sources with $S_{450} > 10$\,mJy.}
\end{figure*}
 
\section{Conclusion - the nature of sub-millimetre galaxies}
From the available evidence I conclude that 
the archetypal ``8-mJy'' sub-millimetre galaxy {\bf i)} lies at $z = 2 -3$, 
{\bf ii)} is forming stars at $\simeq 700\,{\rm M_{\odot}\,yr^{-1}}$, 
{\bf iii)} has a stellar mass of $M_* \simeq 2-3 \times 
10^{11}\,{\rm M_{\odot}}$, {\bf iv)} has a molecular gas mass 
of $M_{\rm gas} 
\simeq 0.5 - 2 \times 10^{11}\,{\rm M_{\odot}}$, {\bf v)} has a 
dynamical mass of $M_{\rm dyn} 
\simeq 3-5 \times 10^{11}\,{\rm M_{\odot}}$, 
{\bf vi)} has an implied dark matter halo mass of $M_h \sim 5 \times 
10^{12}\,{\rm M_{\odot}}$, {\bf vii)} is a fairly mature (possibly interacting)
star-forming disc galaxy with $r_{0.5} \simeq 3\,{\rm kpc}$, and {\bf viii)}
has a SSFR $\simeq 2 - 3 \,{\rm Gyr^{-1}}$, 
as ``expected'' for a normal 
star-forming galaxy at these redshifts (Daddi et al. 2007; Kajisawa et al. 2010; see also the 
value of SSFR = 2.2\,Gyr$^{-1}$ 
for sub-millimetre galaxies at $z \simeq 2.3$ 
reported by Ricciardelli et al. 2010).

Over the next 3 years it will be possible to explore the rest-frame optical 
morphologies of significant samples of sub-millimetre galaxies as a result 
of large-scale {\it HST} WFC3/IR imaging programmes, such as the CANDELS 
survey (Grogin et al. 2011). In addition, the deep far-infrared imaging 
provided by the {\it Herschel} SPIRE+PACS imaging programmes (e.g. HerMES; 
Oliver et al. 2010) has the capablity to provide vastly improved constraints
on dust temperatures, far-infrared luminosities, and hence inferred 
star-formation rates. However, there can be little doubt that the main 
obstacle to obtaining a clean and comprehensive view of the sub-millimetre 
galaxy population remains the lack of unconfused sub-millimetre imaging over
significant areas of sky. This is especially true at the highest redshifts, 
where radio identifications become extremely challenging even with the EVLA, 
and where {\it Herschel} detections can become confined to the 500${\rm \mu m}$
band where the angular resolution delivered is no better than that delivered
by BLAST at 250${\rm \mu m}$ (see Figure 1). As has long been known, what 
is still really required to unlock the power of both the {\it Herschel} 
imaging, and the other key {\it Spitzer} and {\it HST} survey programmes, is 
deep 450${\rm \mu m}$ imaging with SCUBA2 which, mounted on the 15-m JCMT, 
will deliver imaging over degree-scale areas with a 
beam-size of $\simeq 7.5$\,arcsec (FWHM). The power of such imaging 
over that provided by BLAST 
or {\it Herschel} in their longest wavelength channels is 
illustrated in Figure 4. Armed with such data, we can genuinely 
aspire to a proper 
understanding of the role of the sub-millimetre galaxy population in the 
overall story of galaxy formation and evolution.

\acknowledgements 
JSD acknowledges the support of the Royal Society via a Wolfson Research Merit award, and also the support of the European Research Council via the award of an Advanced Grant.
He would also like to acknowledge the work of many collaborators, 
but in particular, for this present paper, 
the contributions of Tom Targett, Michal Michalowski,
Ross McLure, Ed Chapin, Mark Devlin, Dave Hughes, and Michele Cirasuolo.  
This work is based in part on observations made with the NASA/ESA {\it Hubble Space Telescope}, which is operated by the Association 
of Universities for Research in Astronomy, Inc, under NASA contract NAS5-26555.
This work is also based in part on observations made with the {\it Spitzer Space Telescope}, which is operated by the Jet Propulsion Laboratory, 
California Institute of Technology under NASA contract 1407.

\end{document}